\documentclass[10pt, conference, letterpaper]{IEEEtran}
\IEEEoverridecommandlockouts
\usepackage{cite}
\usepackage{algorithm}
\usepackage{amsmath,amssymb,amsfonts}
\usepackage{algorithmic}
\usepackage{graphicx}
\usepackage{subfigure}
\usepackage{textcomp}
\usepackage{enumitem}
\usepackage{xcolor}
\usepackage{fancyhdr}

\def\BibTeX{{\rm B\kern-.05em{\sc i\kern-.025em b}\kern-.08em
    T\kern-.1667em\lower.7ex\hbox{E}\kern-.125emX}}
\begin{document}

\title{SurveilEdge: Real-time Video Query based on Collaborative Cloud-Edge Deep Learning}

\author{
    \IEEEauthorblockN{
        Shibo Wang{${^\dagger}$}, Shusen Yang{${^\ast}$}, Cong Zhao{${^\S}$}
        }
    \IEEEauthorblockA{
        {${^\dagger}$}{${^\ast}$}National Engineering Laboratory for Big Data Analytics, Xi'an Jiaotong University \\
        {${^\dagger}$}Email: wshb20081996@stu.xjtu.edu.cn  {${^\ast}$}Email: shusenyang@mail.xjtu.edu.cn \\
        {${^\S}$}Department of Computing, Imperial College London. Email: c.zhao@imperial.ac.uk
        }
    \thanks{${^\ast}$The corresponding author is Shusen Yang.}
}


\maketitle


\begin{abstract}
The real-time query of massive surveillance video data plays a fundamental role in various smart urban applications
such as public safety and intelligent transportation.
Traditional cloud-based approaches are not applicable because of high transmission latency and prohibitive bandwidth cost, 
while edge devices are often incapable of executing complex vision algorithms with low latency and high accuracy due to restricted resources.
Given the infeasibility of both cloud-only and edge-only solutions, we present SurveilEdge, 
a collaborative cloud-edge system for real-time queries of large-scale surveillance video streams.
Specifically, we design a convolutional neural network (CNN) training scheme to reduce the training time with high accuracy, 
and an intelligent task allocator to balance the load among different computing nodes 
and to achieve the latency-accuracy tradeoff for real-time queries.
We implement SurveilEdge on a prototype\footnote{The code is available at https://github.com/SurveilEdge/SurveilEdge.} with multiple edge devices and a public Cloud, and conduct extensive experiments using real-world surveillance video datasets.
Evaluation results demonstrate that SurveilEdge manages to achieve up to 7$\times$ less bandwidth cost and 
5.4$\times$ faster query response time than the cloud-only solution; 
and can improve query accuracy by up to 43.9\% and achieve 15.8$\times$ speedup respectively, in comparison with edge-only approaches.
\end{abstract}
\section{Introduction}
With the widespread deployment of surveillance cameras, the amount of surveillance video data has been increasing explosively.
The IHS's 2019 report forecasts that over 180 million surveillance cameras will be shipped globally \cite{IHS2019}.
Since it is prohibitively costly and slow to extract information from consecutive surveillance videos with traditional manual manners, 
the automatic and intelligent video analytics has been attracting great attentions from both industry and academia 
\cite{kang2017noscope}\cite{222587}\cite{ananthanarayanan2017real}.

\begin{figure}[htbp]
\centerline{\includegraphics[width=7cm]{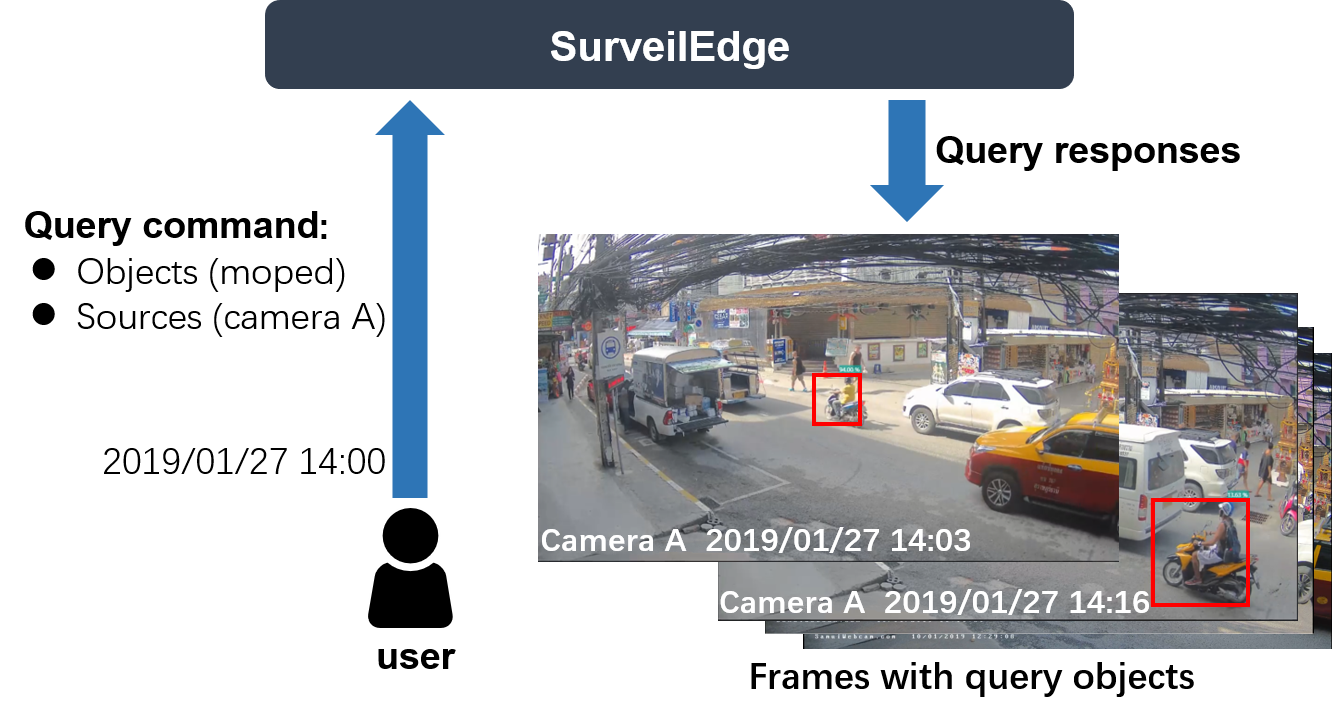}}
\caption{An example of the real-time surveillance video query. Users input the query command includes query objects (moped in this example) and query locations (represented by camera IDs). The system outputs the video frames with query objects in real time.}
\label{example}
\end{figure}

An important application of video analytics is the real-time query of objects (e.g. car, people) within surveillance video streams, which are fundamental in various smart city applications such as event detection\cite{su2011smart}, target tracking\cite{DBLP:journals/corr/WangS15e}, and intelligent traffic management \cite{kumar2005framework}\cite{tang2017vehicle}.
As illustrated in Fig.~\ref{example}, a query task requires real-time responses of video frames containing the objects defined by the users (termed as query objects in the rest of the paper). This requires to process each video frame immediately when receiving it, which is different from previous video query applications (e.g., NoScope \cite{kang2017noscope} and Focus \cite{222587}) that concentrate on queries of recorded video databases.

\par

The development of deep learning (especially CNNs \cite{lecun2015deep,krizhevsky2012imagenet, ren2015faster,he2016deep,he2017mask}) has been promoting a breakthrough of video query applications \cite{jiang2018chameleon,alipourfard2017cherrypick,zhang2017live,venkataraman2016ernest}, which are normally deployed on the Cloud to execute computation-intensive CNN algorithms (e.g. ResNet-152 \cite{he2016deep} and Yolov3\cite{yolov3}). Although cloud-based solutions can guarantee high accuracy by using complex CNN models,  they suffer from increasingly unaffordable bandwidth cost and severe query latency due to the transmission of explosively growing surveillance video data.

Alternatively, the emerging edge computing paradigm \cite{ananthanarayanan2017real}\cite{satyanarayanan2017emergence} enables computation at edge devices, which avoids raw video data transmission to the Cloud.  However, most edge devices are resource-limited where only lightweight CNN models can be deployed (e.g., MobileNet \cite{DBLP:journals/corr/HowardZCKWWAA17}). Therefore, edge-based solutions cannot guarantee trustworthy query accuracy, especially considering that recalls are more crucial in practice.

\begin{figure*}[htbp]
\centerline{\includegraphics[width=18cm]{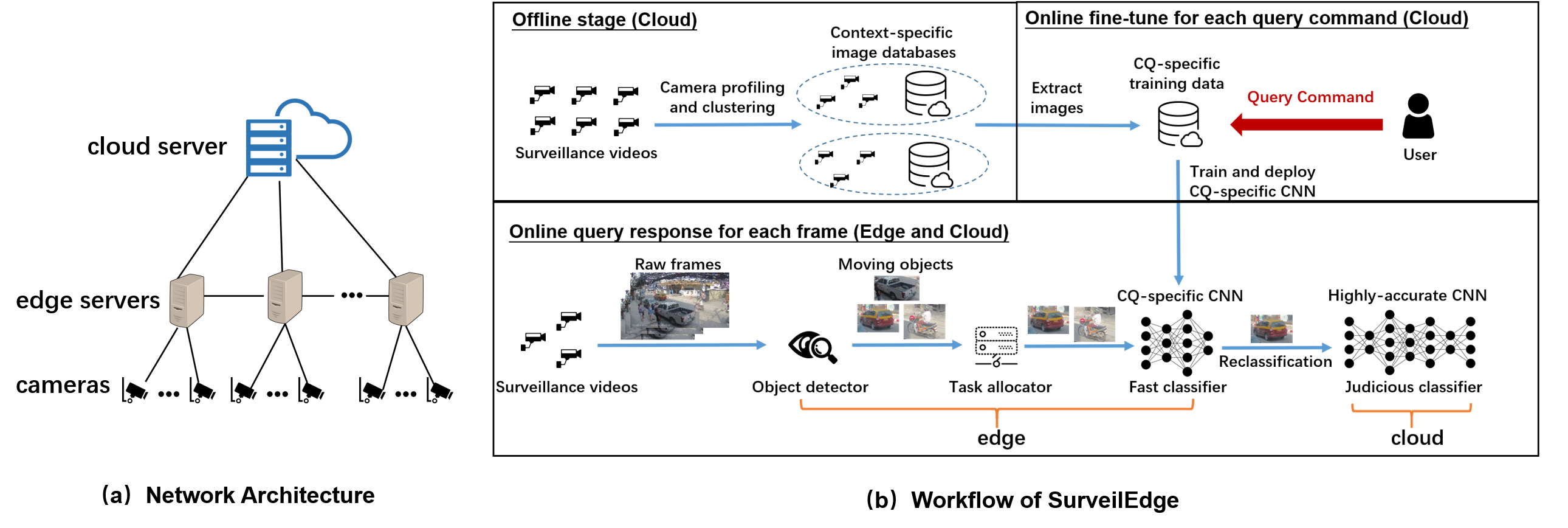}}
\caption{The network architecture and workflow of SurveilEdge}
\label{system}
\end{figure*}

Leveraging advantages of cloud computing and edge computing collaboratively, we present SurveilEdge for real-time queries of large-scale surveillance video streams.
Specifically, we build SurveilEdge and collect over 170 hours of real-world surveillance videos from 14 cameras for performance evaluation.
According to experimental results, SurveilEdge manages to achieve up to 7$\times$ fewer bandwidth cost and 5.4$\times$ faster query response time than the cloud-only solution; and improves query accuracy by up to 43.9\% and achieve 15.8$\times$ speedup respectively, in comparison with the edge-only approach. Our main contributions are threefold.

\begin{itemize}[leftmargin=*]
\item We combine the characteristics of surveillance videos we observed with the fine-tuning technique to train specific CNNs, which significantly reduces the training time while guaranteeing high accuracy.
\item We design an intelligent task allocator with the task scheduling and parameter adjustment algorithm, which not only effectively balances loads among different computing nodes but also achieves the accuracy-latency tradeoff.
\item We conducted extensive experiments based on a real-world prototype and 170-hour surveillance videos. Evaluation results demonstrate that the adaptive cloud-edge SurveilEdge significantly outperforms cloud-only, edge-only, and fixed cloud-edge approaches in terms of query accuracy, bandwidth cost, and query latency.
\end{itemize}





\section{Related Work}




Existing systems on video analytics, such as Chameleon \cite{jiang2018chameleon}, Cherrypick \cite{alipourfard2017cherrypick}, VideoStorm\cite{zhang2017live} and VideoEdge \cite{hung2018videoedge}, consider the resource-quality tradeoff and try to find the optimal configuration for delay-sensitive video analytics applications.
However, all of them consider general video analytics without concentrating on query optimizations for surveillance videos. 
Current video query systems, including Noscope \cite{kang2017noscope}, Focus \cite{222587}, and Blazeit\cite{kang2018blazeit}, conduct specialized optimizations for large-scale video queries and improve the query performance significantly using neural networks.
However, these work mainly focus on queries of recorded video datasets rather than real-time queries of video streams as SurveilEdge.

\section{Overview of SurveilEdge}

We first introduce the system conditions, some characteristics we observed from real surveillance videos, which are considered in the design of SurveilEdge. Then, we describe the overall architecture of SurveilEdge.

\subsection{Spatio-temporal Characteristics of Surveillance Videos}
\label{subsec:Spatio-temporal Characteristics}
In practice,  surveillance cameras that observe similar scenes are likely to produce videos with similar objects. For example, cameras facing a highway and in a building will produce videos of vehicles and persons respectively. In addition, cameras also exhibit the periodicity of busy times, which are also different across monitoring scenes. These spatio-temporal characteristics are also observed by previous work on video analytics~\cite{jiang2018chameleon}.




\subsection{SurveilEdge Overview}

We consider a network composed of cameras, edge devices, and a cloud server, as shown in Fig.\ref{system}(a). Here, each edge device can serve multiple cameras, communicate with other edge devices, and communicate with the cloud server.


As shown in Fig. \ref{system}(b), the workflow of SurveilEdge can be divided into offline and online stages.
The offline stage is conducted on the Cloud, where cameras of similar scenes are categorized into the same context-specific clusters, each of which produces a context-specific training dataset. 
At the online stage, when a user inputs a new query command, a CNN with respect to this query will be trained on the Cloud, based on the context-specific training dataset established in the offline stage. 
Therefore, we call this CNN as a context and query specific (CQ-Specific) CNN.
Here, a CQ-specific CNN is a few-classification CNN that forgoes the full generalization of classifying.
However, the model complexity is reduced, and the capability of identifying certain kinds of objects is enhanced.
The training process normally takes less than a minute with a normal Cloud GPU (e.g. NVIDIA Telsa P4). 
The trained CQ-Specific CNN is then deployed at the corresponding edge server.
With the CQ-specific CNN at the edge and a highly-accurate CNN (e.g. ResNet-152) on the Cloud, each video frame will be processed to produce real-time query responses (normally one to several seconds).  
SurveilEdge adopts an intelligent scheme including task scheduling and parameter adjustment for system load balancing and the accuracy-latency tradeoff of queries.

\section{System Design}

This section describes SurveilEdge in detail.

\subsection{Camera Clustering and Training Dataset Establishment}
\label{subsec:offline}
Since the query accuracy of specific CNNs highly depends on the training data, an intuitive approach is to train one specific CNN using images captured by one camera.
However, this will cause prohibitively high training cost and latency.
Alternatively, it is viable to jointly train a specific CNN to serve a cluster of analogous-scene cameras simultaneously.
Obviously, the cooperation among analogous-scene cameras also swells the volume of training data for each specific CNN, which can effectively prevent the overfitting caused by small training sets.

\begin{figure}[!ht]
\centerline{\includegraphics[width=8.8cm]{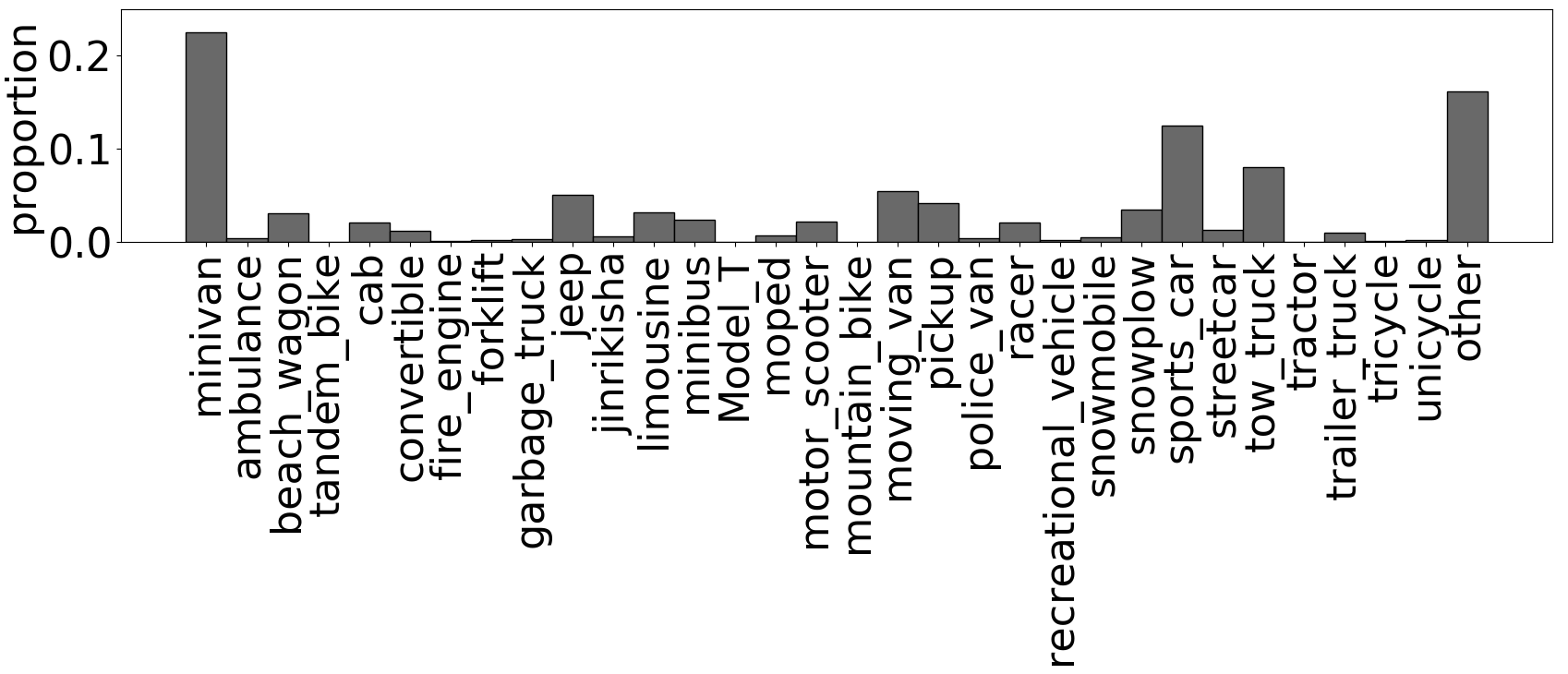}}
\caption{An example of the profile of a camera (i.e., its proportion vector).}
\label{proportion}
\end{figure}

\begin{figure}[!ht]
\centerline{\includegraphics[width=8.8cm]{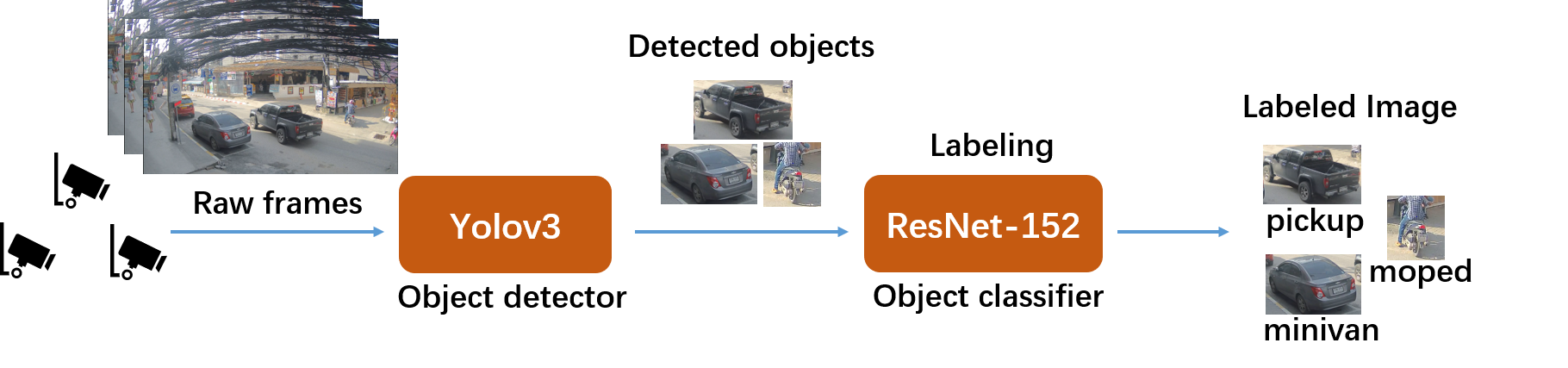}}
\caption{Illustration of training data labeling.}
\label{fig:labelingdata}
\end{figure}

SurveilEdge collects video data at leisure time (without performing query tasks) from each surveillance camera, 
where the collected data are processed using highly-accurate CNNs.
Particularly, Yolov3\cite{yolov3} is used to detect common objects in video streams, and ResNet-152 \cite{he2016deep} is applied to perform more precise and specific classification of detected objects and label them, as shown in Fig. \ref{fig:labelingdata}.

We use the proportion vector to represent the profile of each camera, based on video frames produced by it and uploaded to the Cloud.  Each camera profile is computed as the occurrence frequencies of different objects in all video frames captured by this camera.
An example of a camera profile is shown in Fig.~\ref{proportion}.
The camera profile can roughly represent the scene where the camera is deployed. 
For example, on one hand, if numbers of cars in two surveillance videos are large, but proportions of people are very low, then both cameras tend to be near major roads.
On the other hand, if there are more people and fewer cars, they tend to be near crowded squares or walking trails.

SurveilEdge computes the profile for each camera, and clusters the cameras (based on their profiles) using the K-Means algorithm \cite{hartigan1979algorithm}. Here, the center of a cluster is also a proportion vector, which is regarded as the profile of this cluster.
Cameras within the same cluster are regarded as in the same context and therefore share the same context-specific training dataset. \par

\subsection{Online Fine-tuning of Query-Specific CNNs}

When SurveilEdge receives a new user-defined query command, suitable training data will be selected from the labeled datasets built at the offline stage, using the following method:
Firstly, SurveilEdge randomly selects an appropriate number of labeled images of the query objects as the positive training data.
Then, the negative training data will be selected from labeled images of non-query objects (i.e., objects excluding the user-defined query command).
For a non-query object, more samples will be selected, if its proportion in the cluster profile is larger. This principle of negative training sample selection improves the classification accuracy of commonly observed objects.


Next, SurveilEdge fine-tunes a specific CNN, called CQ-specific CNN, based on pre-trained\footnote{A pre-trained model draws a lot of nutrients from excellent large-scale datasets such as ImageNet and captures universal features like curves and edges.} MobileNet \cite{DBLP:journals/corr/HowardZCKWWAA17} and selected samples above.
Then, fine-tuning with specific video data enhances learning with a higher starting point while maintaining a lower learning rate.
This allows fast convergence of the loss function to achieve outstanding accuracy on relevant query assignments.
For instance, our evaluation shows that it only takes one minute for our fine-tuning scheme to achieve more than 92\% training accuracy with an NVIDIA Tesla P4 GPU.

After the training is accomplished, the weights of specific CNNs are automatically downloaded and deployed on edge devices. \par


\subsection{Moving Object Detection and Query Target Recognition}
As shown in Fig.\ref{system} (b), when the fine-turned CQ-CNN achieves satisfactory accuracy, it will be deployed at corresponding edge devices. Then, SurveilEdge will process video frames to respond to the user-defined query.
Firstly, SurveilEdge determines the presence of moving objects in frames based on the frame difference algorithm. The basic idea of the algorithm is as follows:
the detector extracts consecutive frames from real-time surveillance video streams with a regular time interval (e.g., one second), and checks pixel differences between selected frames.
If pixels within a certain area change apparently across frames, it can be considered  that  a foreground object appears in this area.
Detailed operations are as follows.\par

For a sequence of frames from a video captured by a stationary camera, we extract three consecutive frames $f_{k-1}(x,y,c)$, $f_{k}(x,y,c)$, $f_{k+1}(x,y,c)$ to determine the moving objects in frame $k$, where $c$ represents the $c$th channel of the image.
Then we calculate the per-element absolute differences of successive frames:
\begin{align}
D_1(x,y,c) = | f_{k}(x,y,c) - f_{k-1}(x,y,c) |
\end{align}
\begin{align}
D_2(x,y,c) = | f_{k+1}(x,y,c) - f_{k}(x,y,c) |
\end{align}
Next, we calculate the per-element bit-wise logical conjunction:
\begin{align}
D_a(x,y,c) = D_1(x,y,c) \wedge D_2(x,y,c)
\end{align}
Then we convert the color space of the destination image $D_a(x,y,c)$ into grayscale $D_g(x,y)$, and apply a fixed-level threshold to get the binary image $D_b(x,y)$ for denoising, i.e., filtering out pixels with too small values:

\begin{align}
D_b(x,y) =
\left\{
\begin{array}{lll}
\rm maxval & {\rm if} \ D_g(x,y) > \rm threshold \\
\rm 0 & \rm otherwise \\
\end{array}
\right.
\end{align}
where maxval is the maximum value of corresponding image depth, for example 255 for 8-bit grayscale images. \par

We use some methods of mathematical morphology such as dilation (\ref{dilation}) and erosion (\ref{erosion}), which have low computation complexity, to remove noises and fill the empty of the target:
\begin{align}
D_d(x,y) = \max\limits_{(x',y')\in\Omega_1}D_b(x+x', y+y')
\label{dilation}
\end{align}
\begin{align}
D_e(x,y) = \min\limits_{(x',y')\in\Omega_2}D_d(x+x', y+y')
\label{erosion}
\end{align}
where $(x, y) + \Omega_1$ and $(x, y) + \Omega_2$ represent the neighbourhood of the pixel $(x,y)$.
Finally, we retrieve object bounding boxes from the binary image $D_e(x,y)$ using the algorithm in \cite{suzuki1985topological}. \par

The time interval mentioned above represents the sampling frequency of the query, which is denoted as $s$.
We found that in most scenarios, there is no need to analyze every frames in the video to obtain satisfactory query results.
Besides, SurveilEdge discards some detected images with small sizes or imbalances between length and width to avoid wrong detection results caused by disturbance, and to distinct actual movement from noise.
Compared with object detection algorithms using deep learning, the frame difference algorithm occupies far fewer resources and spends much less time with little decay in accuracy. \par

\begin{figure}[!ht]
\centerline{\includegraphics[width=7cm]{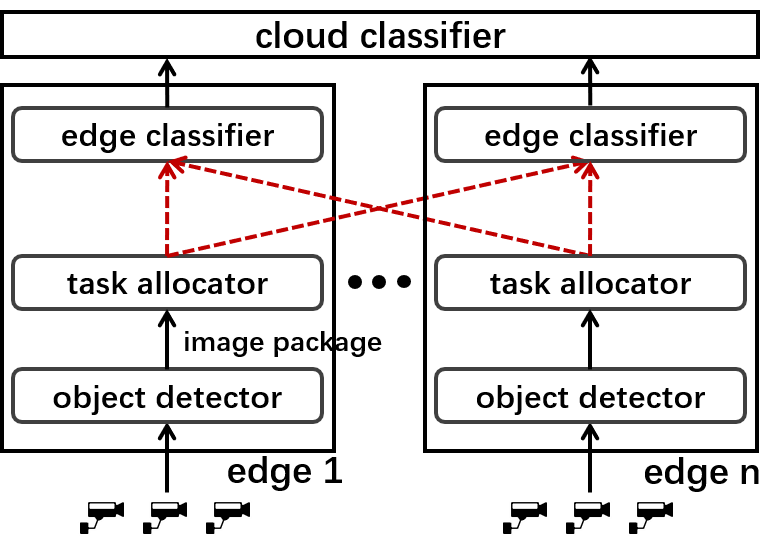}}
\caption{Illustration of task scheduling.}
\label{scheduler}
\end{figure}

The detected foreground images and related information including camera ID, capture time, \emph{etc.} are packaged and passed to the task allocator.
The allocator performs the task scheduling algorithm and determines the destination of each image package (i.e., the package comprised of the foreground image and related information) among all edge devices, as shown in Fig.\ref{scheduler}.
The detailed task scheduling algorithm will be introduced in Subsection \ref{subsec:taskscheduling}. \par

Specific CNNs are deployed in edge classifiers and the high-accuracy CNN is deployed in the cloud classifier.
Each edge classifier maintains a task queue to store image packages waiting to be classified.
When the classification of an image package is finished, the edge classifier gets a new task from the queue and conduct inference if the task queue is not empty. The inference outputs an identification confidence $f$ representing the probability that this image is a query object.
\begin{itemize}
  \item If $f$ is above the upper threshold $\alpha$, this image is confidently considered as a query object.
  \item If $f$ is below the bottom threshold $\beta$, this image is confidently considered as not a query object.
  \item If $f$ is between $\beta$ and $\alpha$, this image cannot be confidently classified by the CQ-CNN at the edge. In this case, this image will be transmitted to the Cloud and be classified with the highly-accurate CNN (e.g. ResNet-152).
\end{itemize}

The setting of thresholds has a pronounced influence on the volume of data uploaded to the Cloud and further affects the computing load of the entire query system.
To balance the computing pressure over time, we design a parameter adjustment algorithm to tune the parameter values according to the system load.
The detailed parameter adjustment algorithm will be introduced in Subsection \uppercase\expandafter{\romannumeral4}-D. \par

\subsection{Task Scheduling and Parameter Adjustment}
\label{subsec:taskscheduling}
Considering spatio-temporal characteristics of surveillance videos we discussed in Subsection \ref{subsec:Spatio-temporal Characteristics}, it is reasonable to balance the query load among edge devices with heterogeneous busy times. Due to the periodicity of the busy time, the dynamic adjustment of threshold parameters is beneficial to determine the tradeoff strategy between the latency and the accuracy of queries at different times.

We have $N$ edge devices and each edge device $i$ maintains a queue of $Q_i$ image packages waiting to be classified by its classifier.  
Let $t_i$  denotes the estimated inference time of an image package at edge device $i$ \footnote{For simplicity, we ignore transmission time between edge devices. However, it is straightforward to model these latencies in SurveilEdge.}.

\subsubsection{Real-time task scheduling}
For an edge device $i$, when a foreground object is detected, the real-time task scheduler will be triggered immediately to determine which device should classify this image package with the least latency, i.e.

\begin{equation}
d = \underset{1\leq i\leq N}{\arg\min}Q_it_i
\end{equation}

Each edge device $i$ keeps a distributed database (SQLite\footnote{https://www.sqlite.org/index.html} in our implementation) alive to store parameters including $\alpha$, $\beta$, as well as $t_i$ and $Q_i$ for all $i$. The update of $t_i$ and $Q_i$ is triggered by the feedback of edge classifiers. Then, the update of $t_i$ or $Q_i$ triggers the immediate update of $\alpha$ and $\beta$. 

\subsubsection{Real-time updates of $\alpha$ and $\beta$}

If the classification latency exceeds the time interval of a query, we shrink the interval length of $[\beta,\alpha]$.
As a result, fewer images will be uploaded to the Cloud and classified again.
Likewise, if the classification latency does not exceed the time interval of the query, the system is capable of reclassifying more images using the highly-accurate CNN on the Cloud to improve the reliability of the query.
Therefore, the interval width of $[\beta,\alpha]$ will be increased, and more images will be transmitted to the Cloud after edge classifications.
Meanwhile, $\alpha$ is limited within the range of $[0.5,1]$, and $\beta$ is limited within the range of $[0,0.5]$. 
Specifically, $\alpha$ is updated as:

\begin{equation}
\alpha_{new}  = \max\big\{\min\{\alpha_{old} - \gamma_1 \times (Q_d \times t_d - s),1\},0.5 \big\}
\end{equation}
where $\gamma_1 \in (0,1)$ is a weighting parameter.  $\beta$ is updated as:

\begin{equation}
\beta_{new} = \gamma_2 \times (1-\alpha_{new} )
\end{equation}
where $\gamma_2$ is within $(0,1)$ to guarantee that the average of $\alpha$ and $\beta$ is lower than $0.5$.

It should be more cautious and wary to eliminate an image than regarding this image as a query object.
In other words, the recall of query is more important than the precision, since that the query object tends to be sparse in surveillance videos, and missing a query object sometimes can lead to terrible results or even ruin the meaning of query tasks in practice.

\subsubsection{Real-time updates of estimated inference latency}

To better predict inference latencies $t_i, 1\leq i\leq N$, we model such latencies as continuous positive random variables and apply the skewed distribution (\emph{i.e.} lognormal distribution) for the estimation based on recently recorded data.
In particular, due to the existence of the minimum latency attributed to the physical limitation, for example the transmission speed limitation determined by the velocity of light, we use the three-parameter lognormal distribution with a support set $x \in(\gamma,+\infty)$ for some $\gamma \geq 0$.
The probability density function of the three-parameter lognormal distribution is
\begin{align*}
f(x ; \mu, \sigma, \gamma)=\frac{1}{(x-\gamma) \sigma \sqrt{2 \pi}} \exp \left\{-\frac{[\ln (x-\gamma)-\mu]^{2}}{2 \sigma^{2}}\right]
\end{align*}
where $0 \leq \gamma < x, -\infty < \mu < \infty, \sigma > 0$, and $\gamma$ represents the minimum latency in theory.
We use $n$ recently recorded latency values $\mathbf{X}=\left\{x_{1}, x_{2}, \ldots x_{n}\right\}$ to estimate the parameters of the three-parameter lognormal distribution. \par

The log-likelihood function is $\ln L(\mu, \sigma, \gamma | \mathbf{X})=$
\begin{align}
- n\ln \sqrt{2 \pi} \sigma&-\sum_{i=1}^{n} \ln \left(x_{i}-\gamma\right)-\frac{\sum\limits_{i=1}^{n}\left(\ln \left(x_{i}-\gamma\right)-\mu\right)^{2}}{2 \sigma^{2}}
\notag
\\& \operatorname{s.t.}\   \min \left\{x_{1},x_{2}, \ldots, x_{n}\right\}>\gamma
\label{loglikelihood}
\end{align}
By computing partial derivatives of the log-likelihood in (\ref{loglikelihood}), we can obtain the local maximum likelihood estimating equations:
\begin{align}
\frac{\partial \ln L}{\partial \mu}=\frac{1}{\sigma^{2}} \sum_{i=1}^{n}\left[\ln \left(x_{i}-\gamma\right)-\mu\right]=0
\label{derivative1}
\end{align}
\begin{align}
\frac{\partial \ln L}{\partial \sigma}=-\frac{n}{\sigma}+\frac{1}{\sigma^{3}} \sum_{i=1}^{n}\left[\ln \left(x_{i}-\gamma\right)-\mu\right]^{2}=0
\label{derivative2}
\end{align}
\begin{align}
\frac{\partial \ln L}{\partial \gamma}=\sum_{i=1}^{n} \frac{1}{x_{i}-\gamma}+\frac{1}{\sigma^{2}} \sum_{i=1}^{n} \frac{\ln \left(x_{i}-\gamma\right)-\mu}{x_{i}-\gamma}=0
\label{derivative3}
\end{align}
Then we express $\hat\mu$, $\hat\sigma^2$ using $\hat{\gamma}$ by solving (\ref{derivative1}) (\ref{derivative2}) (\ref{derivative3}):
\begin{align}
\hat{\mu}=\frac{1}{n}{\sum\limits_{i=1}^{n} \ln \left(x_{i}-\hat{\gamma}\right)},
\label{estimator1}
\end{align}
\begin{align}
\hat{\sigma}^{2}=\frac{1}{n} \sum_{i=1}^{n}\left(\ln \left(x_{i}-\hat{\gamma}\right)-\hat{\mu}\right)^{2}
\label{estimator2}
\end{align}

We replace $\mu, \sigma$ with estimators (\ref{estimator1}) (\ref{estimator2}) in (\ref{derivative3}) and obtain an equation of $\gamma$ (\ref{equationGamma}).
We solve this equation iteratively and obtain the local maximum likelihood estimator of $\gamma$.
Naturally, the estimators of $\mu$ and ${\sigma}^{2}$ can be readily obtained.
\begin{align}
&\left[\sum_{i=1}^{n} \frac{1}{x_{i}-\gamma}\right]\left[\sum_{i=1}^{n} \ln \left(x_{i}-\gamma\right)-\sum_{i=1}^{n}\left(\ln \left(x_{i}-\gamma\right)\right)^{2}+
\notag\right.
\\
&\left.  \frac{1}{n}\left(\sum_{i=1}^{n} \ln \left(x_{i}-\gamma\right)\right)^{2}\right]-n \sum_{i=1}^{n} \frac{\ln \left(x_{i}-\gamma\right)}{x_{i}-\gamma}=0
\label{equationGamma}
\end{align}

Finally, the estimated latency can be regarded as $\mathrm{E}({X})=\hat\gamma+\exp \left(\hat\mu+\frac{\hat\sigma^{2}}{2}\right)$.
 However, using such an approach to estimate the parameters in the three-parameter lognormal distribution, there is a close agreement between $\mathrm{E}({X})$ and $\overline{x}$, which sometimes leads to huge swings of latency for outliers.
 Therefore, in practice, we use the weighted arithmetic mean of  $\mathrm{E}({X})$ and $\mathrm{Median}(X)=\hat\gamma+e^{\hat\mu}$ to predict the values of latency. \par

Due to the obvious complexity of the estimation using the three-parameter lognormal distribution, it can only be used to conduct long-periodic inference time predictions.
Thereby, we design a simple method to estimate the inference time in real time.
At first, we initialize the value of the estimated inference time with an empirical value.
When the database receives feedback from the edge classifier, we use a self-adaptive weighted mean of the old estimated inference time $t_{old}$ and the new inference time $t_{new}$ in the feedback to dynamically update the value of estimated inference time $t$:
\begin{align}
t=\frac{t_{old}^2+t_{new}^2}{(t_{old}+t_{new})^2}\cdot t_{old}+ \frac{2\cdot t_{old}\cdot t_{new}}{(t_{old}+t_{new})^2}\cdot t_{new}
\label{t_update}
\end{align}

When some extreme values or outliers occur and are returned to the parameter database management system on edge devices, the updating method as (\ref{t_update}) can automatically lower the weights of these abnormal values, and thereby reduce their influences on the updated results, which effectively avoids large swings.
Such a prediction method has lower computation complexity than the lognormal method, and can update the estimated inference time more frequently.
At the same time, the estimation method based on the three-parameter lognormal distribution can compensate for the lower reliability of this simple method in longer periods.

\section{Evaluation}
This section presents our evaluation of SurveilEdge.

\subsection{Methodology}
\noindent \textbf{Video Datasets}:
We collected real surveillance video streams from YouTube live for the performance evaluation.
Specifically, more than 170 hours of video streams from 14 cameras with 1080p resolution and 30 fps frame rate were collected, which covered multiple scenarios such as crossroads, schools, and streets.
We selected 130 hours of video data for the training of specific CNNs (with Yolov3 for detection and ResNet-152 for classification as stated in Subsection \uppercase\expandafter{\romannumeral4}-A).
The remaining data were used for evaluation.
By applying K-Means clustering on proportion vectors, surveillance cameras were divided into two clusters, and two training databases with 140,000 and 75,000 images respectively were determined correspondingly for the training of two specific CNNs. \par

\noindent \textbf{Software Tools}:
SurveilEdge was implemented in Python on Ubuntu 16.04.
Docker containers \cite{merkel2014docker} were adopted for the ease of creation and migration of our distributed system.
Opencv \cite{opencv_library} and scikit-image \cite{scikit-image} were used for video streaming simulations and frame difference-based object detection.
We used Keras \cite{chollet2015keras} and Tensorflow backend \cite{abadi2016tensorflow} for the implementation of CNN training and inference, where standard practices (\emph{e.g.} cross validation and early stopping) were also adopted.
For the communication among different nodes, the extremely lightweight publish/subscribe message transport protocol MQTT \cite{light2017mosquitto} was adopted.
The distributed database management system SQLite was applied on edge devices for database access control.
Detailed versions of the adopted tools are listed in Table \ref{versions}. \par

\begin{table}
\begin{center}
\caption{versions of software tools}
\begin{tabular}{|c|c|c|}
\hline
\textbf{Software Tools} & \textbf{Versions} & \textbf{Deploy place} \\
\hline
Ubuntu & 16.04 & Edges and Cloud\\
\hline
Docker & 18.09.7& Edges and Cloud \\
\hline
Python & 3.6.7& Edges and Cloud \\
\hline
OpenCV & 4.0.0.25& Edges \\
\hline
Scikit-image & 0.14.2 & Edges and Cloud \\
\hline
Keras & 2.2.4& Edges and Cloud \\
\hline
Tensorflow & 1.13.1& Edges and Cloud \\
\hline
Mosquitto & 1.6.2 & Edges and Cloud\\
\hline
SQLite & 3.28.0 & Edges\\
\hline

\end{tabular}
\label{versions}
\end{center}
\end{table}

\noindent \textbf{Parameter Settings}:
We applied MobileNet-v2 as the skeleton of specific CNNs deployed on edge devices, and ResNet-152 as the high-accuracy classifier deployed on the Cloud.
Here, the ResNet-152 was treated as the ground-truth CNN for system performance assessment.
The query interval was set as 1s for each surveillance video (\emph{i.e.} sampling once for every 30 frames).
We took moped as an instance of query objects for system testing and evaluation.
Each edge node was set to serve 3-4 surveillance cameras that produce video data and throw them into the query system constantly in real time. \par

\noindent \textbf{Comparatives}:
For a comprehensive evaluation, we compare the performance of SurveilEdge with that of different solutions.
\begin{itemize}
\item \textbf{SurveilEdge(fixed):} The cloud-edge system without the task scheduling and parameter adjustment algorithm. 
For SurveilEdge(fixed), the training scheme and the detection module are the same as SurveilEdge, but all image packages are classified by local edge classifiers at first without offloading, and threshold values remain constant over time, \emph{i.e.} $\alpha=$ 0.8 and $\beta=$ 0.1.
\item \textbf{Edge-only:} For edge-only, the training scheme and the detection module are the same as SurveilEdge, but all image packages are classified by local edge classifiers based on CQ-specific CNN without task scheduling and cloud classifier.
\item\textbf{Cloud-only:} For cloud-only, the detection module is the same as SurveilEdge, but all image packages are transmitted to the Cloud and classified by the highly-accurate CNN.
\end{itemize}

\noindent \textbf{Performance Metrics}:
We used the accuracy, the average latency and the bandwidth cost to evaluate the performance of aforementioned query schemes including SurveilEdge, SurveilEdge(fixed), the edge-only system and the cloud-only system.
Besides these holistic metrics, we also logged the query latency of each frame from surveillance video streams to assess the variance of query latency. \par

We used the F-score to measure the accuracy of queries, which considers both the precision $p$ and the recall $r$ of queries.
The value of $F_\lambda$ is calculated as:
\begin{align*}
F_\lambda = (1+\lambda^2) \cdot \frac{p\cdot r}{\lambda^2\cdot p+r}
\label{Fvalue}
\end{align*}
In particular, we set $\lambda=2$ and used the $F_2$ measure which emphasizes more on false negatives to assess the accuracy of our prototype, because recall is more crucial than precision for surveillance video query tasks \cite{Rijsbergen:1979:IR:539927}. \par

\begin{table}
\begin{center}
\caption{performances of 4 query schemes for the single edge and cloud setting}
\begin{tabular}{|c|c|c|c|}
\hline
\textbf{} & \textbf{accuracy} & \textbf{average latency} & \textbf{bandwidth cost} \\
\hline
\textbf{SurveilEdge(fixed)} & 81.56\% & 2.120s & 563.5 MB
\\
\hline
\textbf{SurveilEdge} &  88.42\% & 1.018s & 1129.5 MB
\\
\hline
\textbf{edge-only} & 69.03\% & 2.091s & 0 MB
\\
\hline
\textbf{cloud-only}& 100\% & 14.823s & 3400.3 MB
\\
\hline

\end{tabular}
\label{results_sing_table}
\end{center}
\end{table}

\begin{figure*}[htbp]
\centering
 
\subfigure[The PDFs of per frame latency.]{
\begin{minipage}[t]{.48\linewidth}
\centering
\includegraphics[width=8.6cm]{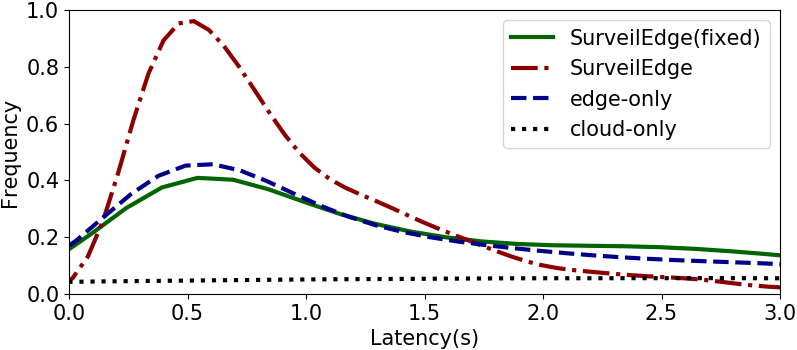}
\end{minipage}%
}%
\subfigure[The line plots of per frame latency (sampled).]{
\begin{minipage}[t]{0.48\linewidth}
\centering
\includegraphics[width=8.6cm]{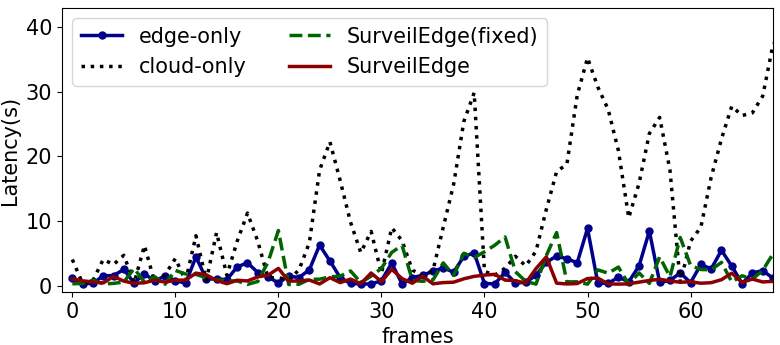}
\end{minipage}%
}
 
\centering
\caption{The performances of four different query schemes for single edge and cloud. }
\label{results_sing}
\end{figure*}


\subsection{Performance of Single Edge and Cloud}
For preliminary tests, we used Docker containers to simulate the edge node and the Cloud on a server with two Intel Xeon E5-2650v4 CPUs and 512 GB DDR4 RAM.
A system with a single edge node and the Cloud was firstly constructed as our primary prototype.

Table.~\ref{results_sing_table} demonstrates experimental results under the single edge and cloud setting.
As shown in Table.~\ref{results_sing_table}, compared with the traditional cloud-only query system, SurveilEdge achieved 3$\times$ fewer bandwidth cost, 14.56$\times$ faster query response, and maintained an acceptable query accuracy.
Compared with the edge-only query system, SurveilEdge improved the query accuracy by 27.5\% and achieved 2.05$\times$ query speedup.
The cloud-edge collaborative architecture of SurveilEdge made a good tradeoff between the accuracy and the query latency.
The classifying and filtering for images on the spot in edge devices reduced the amount of data uploaded to the Cloud and reduced lots of transmission latencies.
Meanwhile, the reclassifying mechanism of uploading some images with weak confidences to the Cloud, compensated the accuracy degradation resulted from classifications using lightweight CNNs in edge devices.

Also, we recorded the query latency of each frame from surveillance video streams.
The Probability Density Functions (PDFs) are shown in Fig.~\ref{results_sing} (a), and the line plots are shown in Fig.~\ref{results_sing} (b).
SurveilEdge not only prominently reduced the average latency of the query but more importantly reduced the variance of the query latency.
Actually, when the query latency of one frame is too large, the query result of this frame will be out of date and meaningless.
The dynamic adjustment of threshold parameters ($\alpha$ and $\beta$) enabled more doubtful images to be identified by highly-accurate classifier during the non-busy time and reduced the uploading volume of data when the whole system was overloaded. It adaptively balanced the system load and the performance of the query.


\begin{table}
\begin{center}
\caption{performances of 4 query schemes for the homogeneous edges and cloud setting}
\begin{tabular}{|c|c|c|c|}
\hline
\textbf{} & \textbf{accuracy} & \textbf{average latency} & \textbf{bandwidth cost} \\
\hline
\textbf{SurveilEdge(fixed)} & 79.36\% & 52.295s & 574.3 MB
\\
\hline
\textbf{SurveilEdge} &  90.7\% & 3.227s & 928.9 MB
\\
\hline
\textbf{edge-only} & 63.55\% & 51.081s & 0 MB
\\
\hline
\textbf{cloud-only}& 100\% & 17.698s & 3066.9 MB
\\
\hline

\end{tabular}
\label{results_muti_table}
\end{center}
\end{table}

\begin{figure*}[htbp]
\centering
 
\subfigure[The PDFs of per frame latency.]{
\begin{minipage}[t]{0.48\linewidth}
\centering
\includegraphics[width=8.6cm]{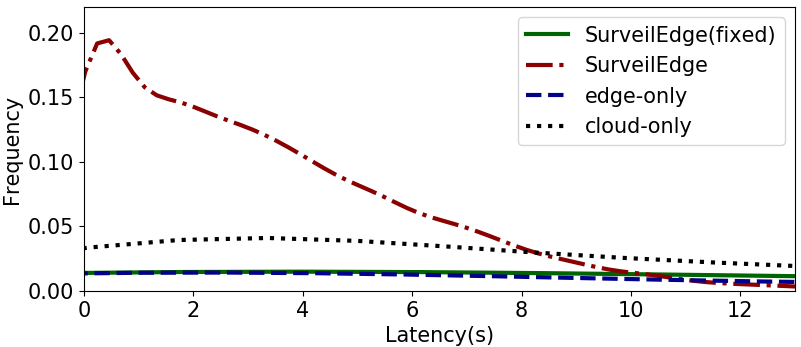}
\end{minipage}%
}%
\subfigure[The line plots of per frame latency on edge device 1 (sampled).]{
\begin{minipage}[t]{0.48\linewidth}
\centering
\includegraphics[width=8.6cm]{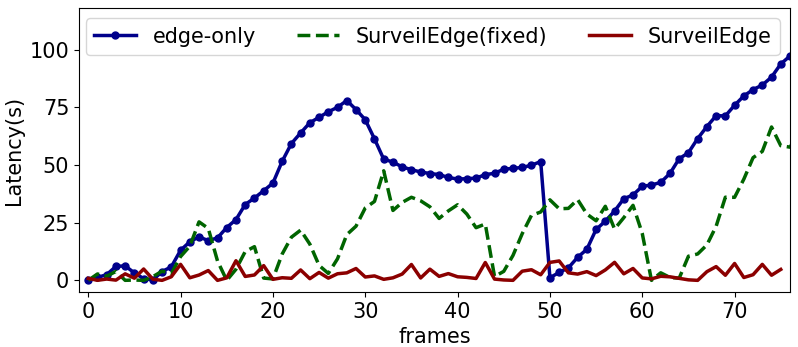}
\end{minipage}%
}%
                 
\subfigure[The line plots of per frame latency on edge device 2 (sampled).]{
\begin{minipage}[t]{0.48\linewidth}
\centering
\includegraphics[width=8.6cm]{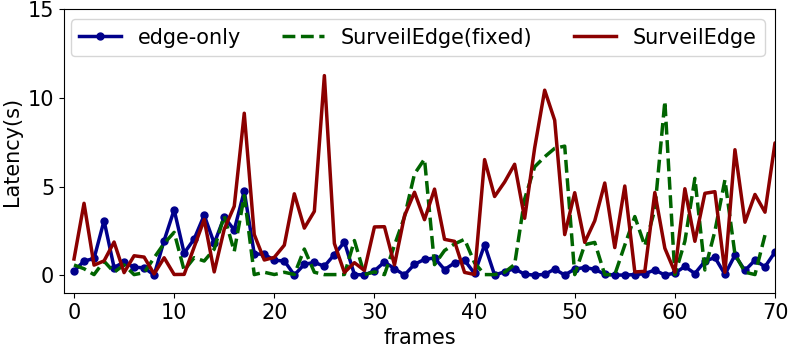}
\end{minipage}
}%
\subfigure[The line plots of per frame latency on edge device 3 (sampled).]{
\begin{minipage}[t]{0.48\linewidth}
\centering
\includegraphics[width=8.6cm]{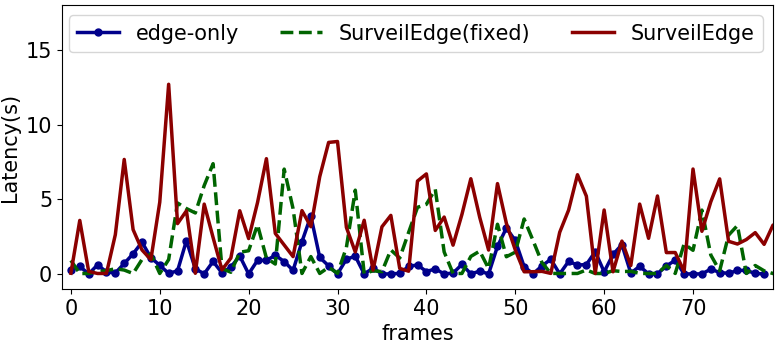}
\end{minipage}
}%
 
\centering
\caption{The performances of four different query schemes for homogeneous edges and cloud. }
\label{results_muti}
\end{figure*}


\subsection{Performance of Homogeneous Edges and Cloud}
Considering the satisfactory simulating results under the single edge and cloud setting, we constructed a real-world prototype with three homogeneous edge devices (each with an Intel Core i7-6700 3.4GHz quad-core CPU and 16 GB DDR4 RAM) and a public Cloud (with 8 logical cores of Intel Xeon E5-2682v4 CPU, 64 GB DDR4 RAM, and an NVIDIA Tesla P4 GPU) to evaluate the performance of SurveilEdge.

Likewise, the accuracy, the average latency and the bandwidth consumption of different query schemes under the homogeneous edges and cloud setting are shown in Table~\ref{results_muti_table}.
Compared with the cloud-only query system, SurveilEdge achieved 3.3$\times$ fewer bandwidth cost, 5.48$\times$ faster query response with an acceptable query accuracy, which proved the advantages of cloud-edge collaborative architecture in SurveilEdge again.

Different from the single edge and cloud setting, the resources of edge devices in our prototype under the homogeneous edges and cloud setting were highly fewer than the resources of the Cloud.
Correspondingly, we can see that SurveilEdge outperformed the edge-only query system and SurveilEdge(fixed)  (15.83$\times$ and 16.2$\times$ speedup, respectively) more obviously compared with the single edge and cloud setting. The comparison with the SurveilEdge(fixed) also implied the indispensable function of the task allocator in SurveilEdge.

We selected a period of busy time for edge device 1 to further show the effects of the task allocator and drew the line plots of per frame latency on three edge devices in Fig.~\ref{results_muti} (b)-(d).
We found that the more limited resources of the edge devices made overloading more frequently especially for edge device 1 in the busy time.
The accumulation of query tasks led to ever-increasing queuing time and skyscraping query latency in the edge-only system and SurveilEdge(fixed).

The intelligent task allocator in SurveilEdge offloaded among the edge devices with different busy times.
When one computing node was overloaded, the task allocator would assign the next task to the other node with relatively less computing pressure.
Therefore, by task scheduling, SurveilEdge managed to prevent excessive and anomalous query latency of certain frames, and significantly reduced the average latency of query compared with the edge-only system and SurveilEdge(fixed).


\begin{figure*}[htbp]
\centering
 
\subfigure[The PDFs of per frame latency.]{
\begin{minipage}[t]{0.48\linewidth}
\centering
\includegraphics[width=8.6cm]{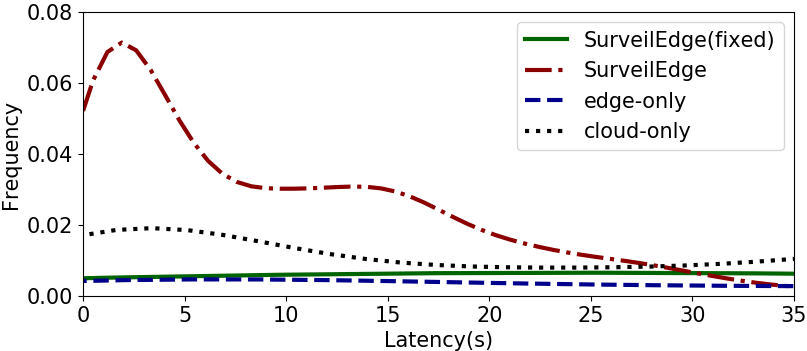}
\end{minipage}%
}%
\subfigure[The line plots of per frame latency on edge device 1 (sampled).]{
\begin{minipage}[t]{0.48\linewidth}
\centering
\includegraphics[width=8.6cm]{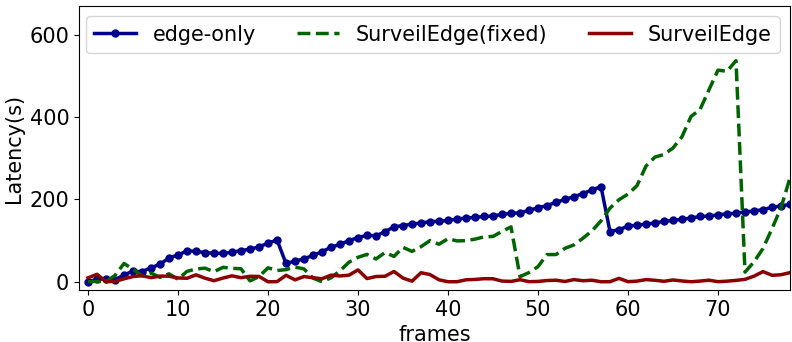}
\end{minipage}%
}%
                 
\subfigure[The line plots of per frame latency on edge device 2 (sampled).]{
\begin{minipage}[t]{0.48\linewidth}
\centering
\includegraphics[width=8.6cm]{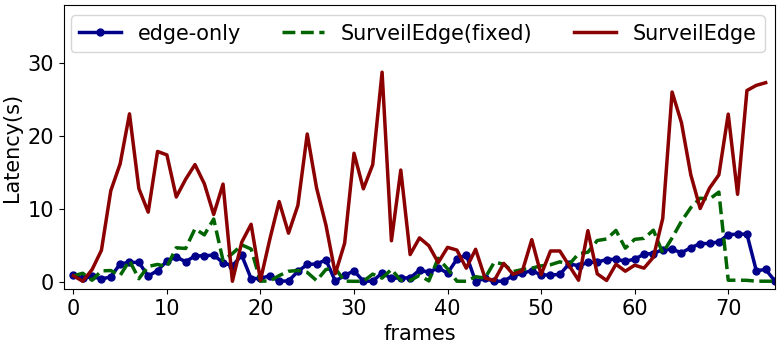}
\end{minipage}
}%
\subfigure[The line plots of per frame latency on edge device 3 (sampled).]{
\begin{minipage}[t]{0.48\linewidth}
\centering
\includegraphics[width=8.6cm]{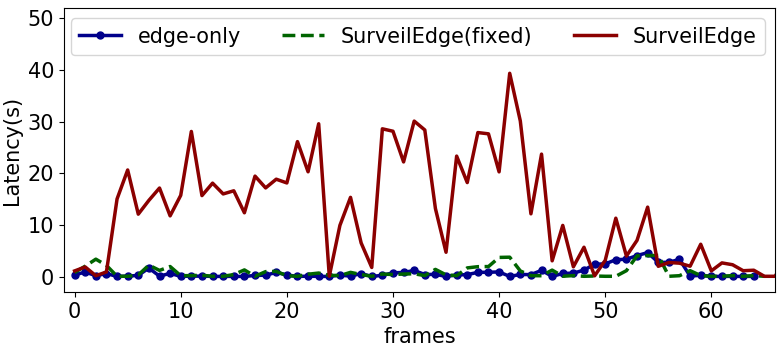}
\end{minipage}
}%
 
\centering
\caption{The performances of four different query schemes for heterogeneous edges and cloud. }
\label{results_hete}
\end{figure*}


\subsection{Performance of Heterogeneous Edges and Cloud}
Considering the resource heterogeneity of practical edge devices, we evaluated the performance of SurveilEdge under the heterogeneous edges and cloud setting.
With the same system components as the homogeneous edges and cloud setting, the resource heterogeneity was introduced by respectively limiting the number of logical CPU cores of each edge device, \emph{i.e.} 2, 4 and 8 cores, using the resource virtualization technique of Docker.

According to the detailed results demonstrated in Table~\ref{results_hete_table}, it is obvious that SurveilEdge outperformed the cloud-only query system in terms of both the average query latency (3.62$\times$ speedup) and the bandwidth consumption (2.9$\times$ fewer).
Compared with the edge-only system and SurveilEdge(fixed), SurveilEdge achieved 10.91$\times$, 11.35$\times$ speedups of query responses, respectively, and improved the accuracy by 22.4\% and 12.3\%, respectively.

\begin{table}
\begin{center}
\caption{performances of 4 query schemes for the heterogeneous edges and cloud setting}
\begin{tabular}{|c|c|c|c|}
\hline
\textbf{} & \textbf{accuracy} & \textbf{average latency} & \textbf{bandwidth cost} \\
\hline
\textbf{SurveilEdge(fixed)} & 75.3\% & 117.23s & 609.2 MB
\\
\hline
\textbf{SurveilEdge} &  86.9\% & 10.33s & 1100.5 MB
\\
\hline
\textbf{edge-only} & 67.5\% & 112.79s & 0 MB
\\
\hline
\textbf{cloud-only}& 100\% & 37.42s & 3197.7 MB
\\
\hline

\end{tabular}
\label{results_hete_table}
\end{center}
\end{table}

From Fig.~\ref{results_hete} (b)-(d), we can see that the resource heterogeneity of three edge devices raised the variance of query latency, larger than that under the homogeneous edges and cloud setting, in the edge-only system and SurveilEdge(fixed).
Edge devices 2 and 3 had relatively more resources and fewer burdens of query tasks, which resulted in less query latency.
However, the lack of computing power on edge device 1 made the accumulation of tasks worse and brought larger waiting latency.

According to Fig.~\ref{results_hete} (b)-(d), we find that in SurveilEdge, edge devices 2 and 3 shared query tasks of edge device 1 under the control of the task allocator.
The cooperation among edge devices with different busy times improved the resource utilization of the entire query system and avoided excessive queuing time. Therefore, both the mean value and variance of query latency were reduced prominently with SurveilEdge.
 
\section{Conclusion and Future Works}
In this paper, we present a cloud-edge collaborative and real-time query system, SurveilEdge, for large-scale surveillance video streams, which significantly reduces the query latency and the bandwidth cost while guaranteeing an acceptable accuracy.
Moreover, the intelligent task allocator with the task scheduling and parameter adjustment algorithm manages to balance the system load and reduce the variance of per frame query latency.
We implement SurveilEdge on the real-world prototype with a public Cloud and multiple homogeneous/heterogeneous edge devices for the performance evaluation based on real surveillance videos.
Experimental results demonstrate the effectiveness and advantages of our approach.
In fact, our schemes and system architecture are not limited to the particular use case in this paper, and they can be beneficial to most latency-sensitive applications based on deep learning.
We will be pursuing a better tradeoff among latency, bandwidth cost and accuracy for broadened latency-sensitive applications in the future.

\section*{Acknowledgement}
We would like to thank the anonymous reviewers for their thoughtful suggestions. This work was supported in part by the National Natural Science Foundation of China under Grant 61772410, Grant 61802298, and Grant U1811461, in part by National Key Research and Development Program of China under Grant 2017YFB1010004, in part by the China 1000 Young Talents Program, and in part by the Young Talent Support Plan of Xi'an Jiaotong University.

\bibliographystyle{IEEEtran.bst}
\bibliography{IEEEabrv.bib,Bibtex.bib}

\end{document}